# Quasistatic limit of the strong-field approximation describing atoms in intense laser fields: Circular polarization

Jarosław H. Bauer *

*Katedra Fizyki Teoretycznej Uniwersytetu Łódzkiego,*
*Ul. Pomorska 149/153, PL-90-236 Łódź, Poland*

In the recent work of Vanne and Saenz [Phys. Rev. A **75**, 063403 (2007)] the quasistatic limit of the velocity gauge strong-field approximation describing the ionization rate of atomic or molecular systems exposed to linearly polarized laser fields was derived. It was shown that in the low-frequency limit the ionization rate is proportional to the laser frequency $\omega$ (for a constant intensity of the laser field). In the present work I show that for circularly polarized laser fields the ionization rate is proportional to $\omega^4$ for $H(1s)$ and $H(2s)$ atoms, to $\omega^6$ for $H(2p_x)$ and $H(2p_y)$ atoms, and to $\omega^8$ for $H(2p_z)$ atoms. The new analytical expressions for asymptotic ionization rates (which become accurate in the limit $\omega \to 0$) contain no summations over multiphoton contributions. For very low laser frequencies (optical or infrared), these expressions usually remain with an order-of-magnitude agreement with the velocity gauge strong-field approximation.

___________

*bauer@uni.lodz.pl

The name "the strong-field approximation" (SFA) is frequently used to call one of two well-known versions of the $S$-matrix theory, which describes nonresonant multiphoton ionization of atoms and ions in intense laser fields [1,2]. In principle, this name (SFA) fits well to both theories, where the main approximation is connected with the use of the Gordon-Volkov wave function [3,4], as a final state of an outgoing electron. The basic difference between these two versions of the $S$-matrix theory is the Hamiltonian form of the laser-atom interaction. Keldysh used it in the length gauge (LG) [1], while Reiss used this Hamiltonian in the velocity gauge (VG) [2] in their pioneering works. The common feature of both approaches [1,2] (but also present in numerous later papers) is the application of nonrelativistic and dipole approximations to a description of atoms (or ions) in intense laser fields. It is now obvious that theories using such approximations are not always sufficient for presentday experiments. In superstrong laser fields first nondipole (i.e. connected with a magnetic-field component of an electromagnetic plane wave), and then relativistic effects have to be taken into account [5-8]. Let us note that the magnetic-field component of the strong, but nonrelativistic, laser field is less essential for a circular polarization (CP) than for a linear polarization (LP) of the field [7,8]. This is because a classical free electron in the circularly polarized plane-wave electromagnetic field, even in the fully relativistic regime, moves along a circle lying in the polarization plane (in the simplest frame of reference; see Sec. 48 (p. 134) of Ref. [9] and Ref. [7]). In contrast, in the fully relativistic linearly polarized plane-wave field, the motion takes place along the "figure-8" path in the plane determined by the polarization direction and the propagation direction (in the simplest frame of reference; see Refs. [9,6-8]). However, even in the case of the LP, nonrelativistic nondipole tunneling theories [5,10,11] proved correct in experiments for higher laser intensities than expected [12,13].

In the recent work of Vanne and Saenz [14] the quasistatic limit of the VG SFA in linearly polarized laser fields was derived. It appears, that in the quasistatic limit ($\omega \to 0$) the ionization rate is proportional to the laser frequency. Naturally, the question then arises, how the VG SFA ionization rate behaves in the quasistatic limit, when the laser field is circularly polarized. This is the main purpose of my present work. The expressions, which describe ionization rates in the VG SFA, were given in Ref. [2] [for the $H(1s)$ atom] and in Ref. [15] [for the $H(1s)$, $H(2s)$, $H(2p_x)$, $H(2p_y)$, and $H(2p_z)$ atoms; see Eqs. (A.11)-(A.14) in appendix A therein]. Of course, the nonrelativistic Gordon-Volkov wave function was used to derive the ionization rates in Refs. [2,15]. On the other hand, in the works of Krainov and Shokri [16,17], an improved low-frequency theory was presented. It is based on multiplying the Gordon-Volkov wave function by the quasi-classical Coulomb correction factor of $\left(2Z^2/n^2 Fr\right)^n$ ($Z$ - the nuclear charge, $n$ - the effective principal quantum number of the initial atomic or ionic state, $F$ - the amplitude of the electric field vector of the laser, and $r$ - the distance between the electron and the nucleus). (In the present work I use atomic units: $\hbar = e = m_e = 1$, and I substitute explicitly $-1$ for the electronic charge.). The results of Refs. [16,17] are valid not only when

$$F < F_{BSI} = \frac{Z^3}{16 n^4}, \qquad (1)$$

where $F_{BSI}$ is the so-called barrier-suppression field strength [see, e.g. Eqs. (4) and (5) in Ref. [16]]. $F_{BSI}$ is a kind of a critical field strength above which the atom should, in principle, ionize immediately in the quasistatic limit. Equation (1) gives $F_{BSI} = 0.0625$ for the $H(1s)$ atom. Another critical field strength for the $H(1s)$ atom was given in Refs. [18,19]. Namely, it was $F_{cr} \approx 0.15$, which seems to be a more realistic value. In the quasistatic limit one has $\omega \to 0$. Therefore, for a sufficiently low frequency of the laser, the Keldysh parameter $\gamma$ [1] obeys

$$\gamma = \frac{Z\omega}{nF} \ll 1. \qquad (2)$$

To derive the new analytical (asymptotic) formulas, I will assume that Eqs. (1) and (2) are satisfied, although my numerical results [20] (see also Figs. 1-4 here) indicate that for $F \geq F_{BSI}$ the new formulas are also useful, if $\omega$ is sufficiently low.

The present notation resembles the one from Refs. [16,17]. In this notation Eqs. (A.11)-(A.14) from Ref. [15] take the form

$$\Gamma_{1s} = \sum_{N=N_0}^{\infty} \frac{8Z^5 p_N}{\left(2E_B + p_N^2\right)^2} \int_0^\pi d\vartheta \sin\vartheta J_N^2\left(\frac{p_N F \sin\vartheta}{\omega^2}\right), \qquad (3)$$

$$\Gamma_{2s} = \sum_{N=N_0}^{\infty} \frac{Z^5 p_N \left(-2E_B + p_N^2\right)^2}{\left(2E_B + p_N^2\right)^4} \int_0^\pi d\vartheta \sin\vartheta J_N^2\left(\frac{p_N F \sin\vartheta}{\omega^2}\right), \qquad (4)$$

$$\Gamma_{2p_x} = \Gamma_{2p_y} = \sum_{N=N_0}^{\infty} \frac{Z^7 p_N^3}{2\left(2E_B + p_N^2\right)^4} \int_0^\pi d\vartheta \sin^3\vartheta J_N^2\left(\frac{p_N F \sin\vartheta}{\omega^2}\right), \qquad (5)$$

$$\Gamma_{2p_z} = \sum_{N=N_0}^{\infty} \frac{Z^7 p_N^3}{\left(2E_B + p_N^2\right)^4} \int_0^\pi d\vartheta \sin\vartheta \cos^2\vartheta J_N^2\left(\frac{p_N F \sin\vartheta}{\omega^2}\right), \qquad (6)$$

where $p_N = \left(2N\omega - F^2/\omega^2 - 2E_B\right)^{1/2}$ is the asymptotic momentum of the outgoing electron, $E_B = Z^2/(2n^2)$ is the binding energy of the initial state of the atom, $N$ is the number of photons absorbed, and $N_0 = [z + E_B/\omega] + 1$ is the minimal value of $N$. [The symbol $[x]$ denotes the integer part of the (positive) number $x$. For the laser field intensity $I$, and the CP, the following relation holds: $I = 2F^2 = 4z\omega^3$.] It appears, that in each of Eqs. (3)-(6) the main contribution to the respective sum comes from the terms, where the final (asymptotic) kinetic energy ($E_N = p_N^2/2$) of the outgoing electron is close to the ponderomotive potential $U_P = z\omega = F^2/(2\omega^2)$. For the ordinary Bessel function $J_N$ (from Eqs. (3)-(6)) the following asymptotic expansion will be used

$$J_N\left(\frac{N}{\cosh\alpha}\right) \approx \left(2\pi N \tanh\alpha\right)^{-1/2} \exp[N(\tanh\alpha - \alpha)], \qquad (7)$$

which is valid for $0 < \alpha \ll 1$. One substitutes $N/\cosh\alpha = p_N F \sin\vartheta/\omega^2$. Then, in Eq. (7), one applies the Taylor expansion to the argument of $J_N$ and the argument of an exponential function, retaining every time two nonzero terms of the lowest order. After solving the resulting algebraic equation for the small quantity $\alpha$, one substitutes this $\alpha$ to the exponent in Eq. (7). In the pre-exponential factor the approximation $\tanh\alpha \approx \alpha$ is sufficient. [For more detail see Sec. 4 of Ref. [17].] As a result, one obtains

$$J_N^2\left(\frac{p_N F \sin\vartheta}{\omega^2}\right) \approx \frac{\omega}{2\pi Z p_N} \exp\left[-\frac{2Z^3}{3n^3 F}\left(1 - \frac{\gamma^2}{15}\right) - \frac{ZF}{n\omega^2}\left(\frac{\pi}{2} - \vartheta\right)^2 - \frac{Z\omega^4}{nF^3}\delta N'^2\right], \qquad (8)$$

where $\delta N' = \delta N - F^2/(6n^2\omega)$ and $\delta N = N - F^2/\omega^3 - E_B/\omega$. Let us note that the previous expression (8) depends on the initial-state principal quantum number $n$ (also through $p_N$, $\gamma$, and $\delta N'$). When Eqs. (1) and (2) are satisfied, angular distributions of outgoing electrons are strongly peaked at $\vartheta = \pi/2$, i.e. electrons are mainly emitted in the polarization plane. Equation (8) can be used now in Eqs. (3)-(6), leading to simple gaussian integrals upon $\vartheta' = \pi/2 - \vartheta$, where only a very narrow vicinity of $\vartheta' = 0$ matters. Therefore, in the integrals upon $\vartheta'$, it is enough to take into account only the first nonzero term of the Taylor expansion of trigonometric functions, which appear in the integrands from Eqs. (3)-(6). For the $H(1s)$ atom (one puts $n=1$) the respective integral is

$$\int_0^\pi d\vartheta \sin\vartheta J_N^2\left(\frac{p_N F \sin\vartheta}{\omega^2}\right) \approx \frac{\omega^2}{2\sqrt{\pi Z^3 F p_N}} \exp\left[-\frac{2Z^3}{3F}\left(1-\frac{\gamma^2}{15}\right) - \frac{Z\omega^4}{F^3}\delta N'^2\right]. \quad (9)$$

For $H(2s)$, $H(2p_x)$ (or $H(2p_y)$), and $H(2p_z)$ atoms (one puts $n=2$) the respective integrals are

$$\int_0^\pi d\vartheta \sin\vartheta J_N^2\left(\frac{p_N F \sin\vartheta}{\omega^2}\right) \approx \frac{\omega^2}{\sqrt{2\pi Z^3 F p_N}} \exp\left[-\frac{Z^3}{12F}\left(1-\frac{\gamma^2}{15}\right) - \frac{Z\omega^4}{2F^3}\delta N'^2\right], \quad (10)$$

$$\int_0^\pi d\vartheta \sin^3\vartheta J_N^2\left(\frac{p_N F \sin\vartheta}{\omega^2}\right) \approx \frac{\omega^2}{\sqrt{2\pi Z^3 F p_N}} \exp\left[-\frac{Z^3}{12F}\left(1-\frac{\gamma^2}{15}\right) - \frac{Z\omega^4}{2F^3}\delta N'^2\right], \quad (11)$$

$$\int_0^\pi d\vartheta \sin\vartheta \cos^2\vartheta J_N^2\left(\frac{p_N F \sin\vartheta}{\omega^2}\right) \approx \frac{\omega^4}{\sqrt{2\pi Z^5 F^3 p_N}} \exp\left[-\frac{Z^3}{12F}\left(1-\frac{\gamma^2}{15}\right) - \frac{Z\omega^4}{2F^3}\delta N'^2\right]. \quad (12)$$

Substituting Eqs. (9)-(12) in Eqs. (3)-(6), respectively, one obtains asymptotic ionization rates in the form $\Gamma^{asympt} = \sum_{N=N_0}^\infty \Gamma_N(Z,\omega,F)$, where $\Gamma_N(Z,\omega,F)$ denote partial ionization rates corresponding to absorption of exactly $N$ photons. Successive terms $\Gamma_N(Z,\omega,F)$ of the previous sum form the kinetic energy spectrum of outgoing electrons (if $\Gamma_N$ is shown as a function of $E_N$). When $F = const$ and $\omega \to 0$, one obtains that $N_0 \to \infty$, but one can change the index of summation to $\delta N'$ [as defined right below Eq. (8)]. In the limit $\omega \to 0$ $\delta N'$ changes from $-\infty$ to $+\infty$, but Eqs. (9)-(12) clearly show that the main contribution to each sum comes from terms with $\delta N' \approx 0$. Furthermore, one can transform each sum over $\delta N'$ to a gaussian integral. Since $p_N$ is large and changes much slower with $\delta N'$ than the exponential factor, one can put $p_N^2 \approx F^2/\omega^2 \gg E_B$ and neglect $E_B$ in the resulting integrand. This makes the integration upon $\delta N'$ trivial. Finally, one obtains the following asymptotic expressions describing the VG SFA ionization rates

$$\Gamma_{1s}^{asympt} = \frac{4Z^3\omega^4}{F^3} \exp\left[-\frac{2Z^3}{3F}\left(1-\frac{\gamma^2}{15}\right)\right], \tag{13}$$

$$\Gamma_{2s}^{asympt} = \frac{Z^3\omega^4}{F^3} \exp\left[-\frac{Z^3}{12F}\left(1-\frac{\gamma^2}{15}\right)\right], \tag{14}$$

$$\Gamma_{2p_x}^{asympt} = \Gamma_{2p_y}^{asympt} = \frac{Z^5\omega^6}{2F^5} \exp\left[-\frac{Z^3}{12F}\left(1-\frac{\gamma^2}{15}\right)\right], \tag{15}$$

$$\Gamma_{2p_z}^{asympt} = \frac{Z^4\omega^8}{F^6} \exp\left[-\frac{Z^3}{12F}\left(1-\frac{\gamma^2}{15}\right)\right]. \tag{16}$$

Equations (13)-(16) are the main result of this work. For the $H(1s)$, $H(2s)$, and $H(2p_z)$ atoms their quantum numbers (without spin) $(n,l,m)$ are the following: $(1,0,0)$, $(2,0,0)$, and $(2,1,0)$ respectively. Wave functions of the atoms $H(2p_x)$ and $H(2p_y)$ are linear combinations of wave functions (normalized to unity) with the quantum numbers $(2,1,-1)$ and $(2,1,1)$, namely

$$\Phi_{2p_x} = \frac{1}{\sqrt{2}}\left(\Phi_{2,1,-1} - \Phi_{2,1,1}\right), \qquad \Phi_{2p_y} = \frac{i}{\sqrt{2}}\left(\Phi_{2,1,-1} + \Phi_{2,1,1}\right). \tag{17}$$

Since the wave functions $\Phi_{2,1,-1}$ and $\Phi_{2,1,1}$ are orthogonal, one can show [20] that in the VG SFA (unlike in its LG counterpart) the ionization rate, in the circularly polarized laser field, is the same for all the initial states given by

$$\Phi_{2p} = \alpha \Phi_{2,1,-1} + \beta \Phi_{2,1,1}, \tag{18}$$

where $\alpha$ and $\beta$ are arbitrary complex numbers such that $|\alpha|^2 + |\beta|^2 = 1$. In particular one has

$$\Gamma^{asympt}_{2,1,-1} = \Gamma^{asympt}_{2,1,1} = \Gamma^{asympt}_{2p_x} = \Gamma^{asympt}_{2p_y}. \tag{19}$$

When the electron is initially bound ($E_B = \kappa^2/2$) in the zero-range potential with the initial-state wave function (normalized to unity) given by

$$\Phi(\vec{r}) = \sqrt{\frac{\kappa}{2\pi}} \frac{\exp(-\kappa r)}{r}, \tag{20}$$

the ionization rate for the circularly polarized laser field in the VG SFA is [2]

$$\Gamma_{zero-range} = \sum_{N=N_0}^{\infty} \kappa p_N \int_0^{\pi} d\vartheta \sin\vartheta J_N^2\left(\frac{p_N F \sin\vartheta}{\omega^2}\right). \tag{21}$$

If one utilizes Eq. (9) (with $Z = \kappa$) in Eq. (21), and one makes analogous calculations, one finally obtains

$$\Gamma^{asympt}_{zero-range} = \frac{F}{2\kappa} \exp\left[-\frac{2\kappa^3}{3F}\left(1 - \frac{\gamma^2}{15}\right)\right]. \tag{22}$$

The result (22) is in agreement with the old result for the ionization in a static electric field [21], if one puts $\gamma = 0$ in the last equation (what corresponds to $\omega = 0$). Therefore, for the zero-range binding potential, the ionization rate in the circularly polarized laser field with $\omega \to 0$ is exactly the same, as in the static field of the same amplitude $F$. This is not true in the VG SFA, if the binding potential is the Coulomb one. For example, instead of the well-known [21-23] static-field expression for the $H(1s)$ atom

$$\Gamma_{stat} = \frac{4Z^5}{F} \exp\left(-\frac{2Z^3}{3F}\right),\qquad(23)$$

Eq. (13) for $\gamma = 0$ gives $\Gamma_{1s}^{asympt} \approx (4Z^3\omega^4/F^3)\exp(-2Z^3/3F)$. The latter expression has only the same exponential factor, as the correct expression (23). The proportionality coefficient $\omega^4$, which exists in Eq. (13), is the counterpart (for the CP) of the proportionality coefficient $\omega$ found formerly by Vanne and Saenz in Ref. [14] (for the LP). However, unlike for the LP, the power of $\omega$ in the coefficient depends on the $(n,l,m)$ quantum numbers of the initial state for the CP [cf. Eqs. (13)-(16)]. As a result, for both polarizations of the laser field, $I = const$ and $\omega \to 0$ lead to nulling of the ionization rate. The authors of Ref. [14] state that "This evidently unphysical result indicates a breakdown of the SFA." (see Sec. IV of Ref. [14]) and they propose an application of a Coulomb correction factor (based on the results of Ref. [10]) to the VG SFA ionization rate formula. I will not proceed further in this direction in the present Brief Report for the CP. However, for the $H(1s)$ atom and the VG SFA such Coulomb-corrected theory has been recently proposed for both polarizations [24].

In Figs. 1-4 there are the VG SFA ionization rates as a function of intensity (Figs. 1 and 2) or frequency (Figs. 3 and 4) for the $H(1s)$ atom in the strong circularly polarized laser field. In each plot I compare the exact and the asymptotic results. Solid lines correspond to the exact ones [from Eq. (3)] and dotted lines correspond to the asymptotic ones [from Eq. (13)]. In Figs. 1-4 the field parameters $(\omega, I)$ cover a total range of a validity of the nonrelativistic SFA. In Figs. 1 and 2 the laser frequencies are fixed. Both frequencies are of an experimental interest. $\omega = 0.0043$ $a.u.$ conforms with $CO_2$ laser radiation ($\lambda = 10.6$ μm), and $\omega = 0.057$ $a.u.$ conforms with $Ti:sapphire$ laser radiation ($\lambda = 800$ nm). There are also two vertical lines (in each of Figs. 1 and 2), which show $I_{BSI} = 2F_{BSI}^2$ and $I_{cr} = 2F_{cr}^2$ for the $H(1s)$ atom. The agreement between exact and asymptotic ionization rates, particularly for the lower frequency $\omega = 0.0043$ $a.u.$ is satisfactory. For $\omega = 0.057$ $a.u.$ it is hard to obey both the conditions (1) and (2). Therefore, in the case of Fig. 2, Eq. (13) can be treated rather as a useful approximation (the upper bound) to Eq. (3). Nevertheless, if $F \sim F_{cr}$ (or lower) is fixed, the agreement between exact and asymptotic results is very good for sufficiently low frequencies (see Fig. 3, where $F = 0.02$ $a.u.$). In Fig. 4 $F = 2$ $a.u.$, which is well above $F_{cr}$. For $\omega \to 0$ there are to parallel lines in the log-log plot. This fact indicates that also for $F > F_{cr}$ (in the limit $\omega \to 0$) the VG SFA ionization rate is proportional to $\omega^4$.

In conclusion, I have derived approximate formulas for the VG SFA ionization rate for the hydrogenic atom in the initial state, which is described by the principal quantum number $n$ either equal to 1 or 2. It appears that the respective ionization rate is proportional to $\omega^4$ (for $(n,l,m) = (1,0,0)$ or $(n,l,m) = (2,0,0)$), to $\omega^6$ (for $(n,l,m) = (2,1,\pm 1)$), and to $\omega^8$ (for $(n,l,m) = (2,1,0)$). These asymptotic expressions become exact in the quasistatic limit. For finite, but low frequencies, these expressions may be treated as very simple upper bounds to the exact VG SFA expressions. The latter become much more time-consuming (in numerical calculations) in the infrared or far-infrared frequency regime.

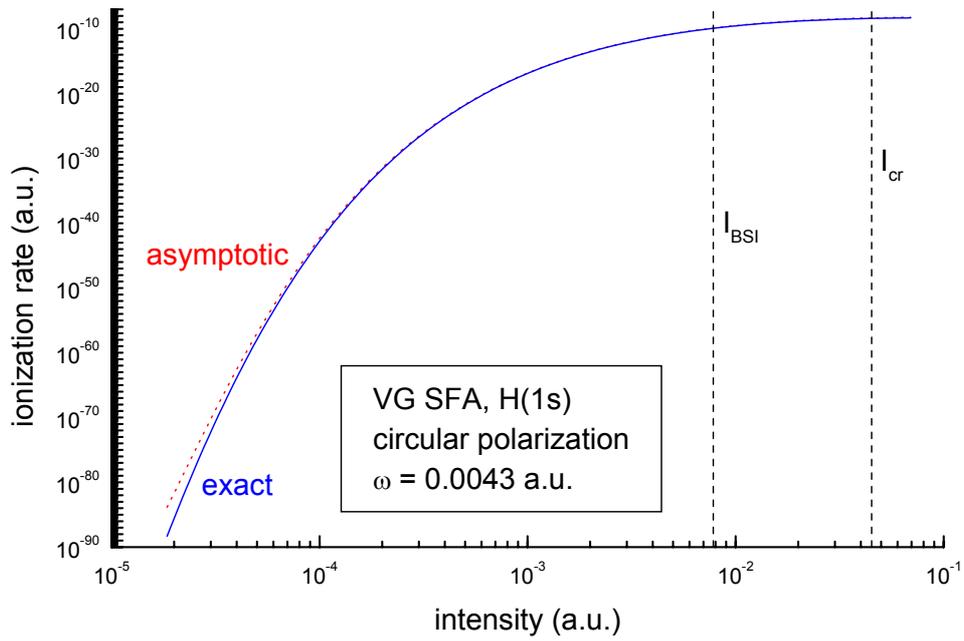

FIG. 1. The VG SFA ionization rates of the $H(1s)$ atom in the circularly polarized laser field for $\omega = 0.0043$ *a.u.* versus intensity of the field (see text for more detail).

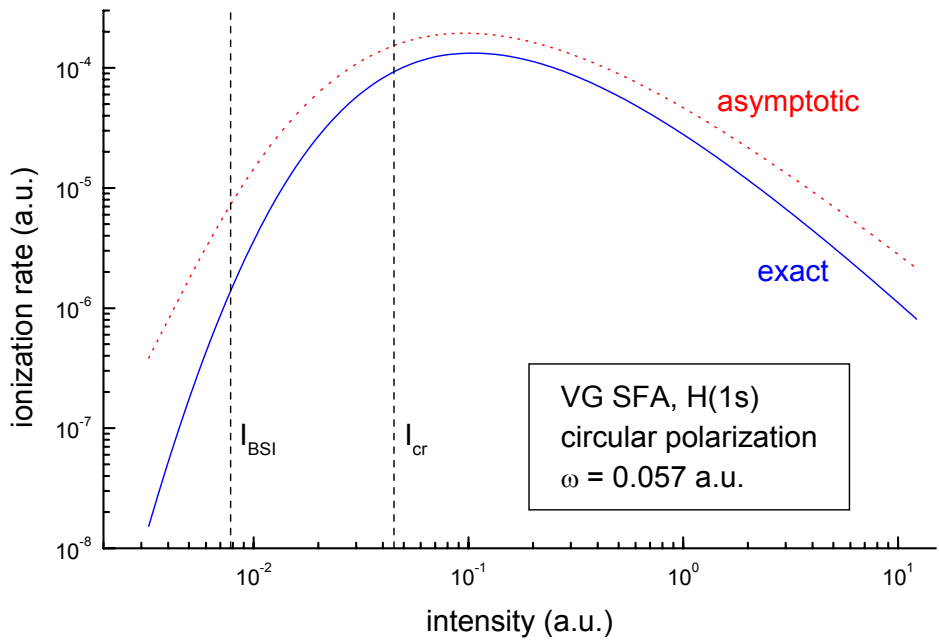

FIG. 2. As Fig. 1, but for $\omega = 0.057$ *a.u.*

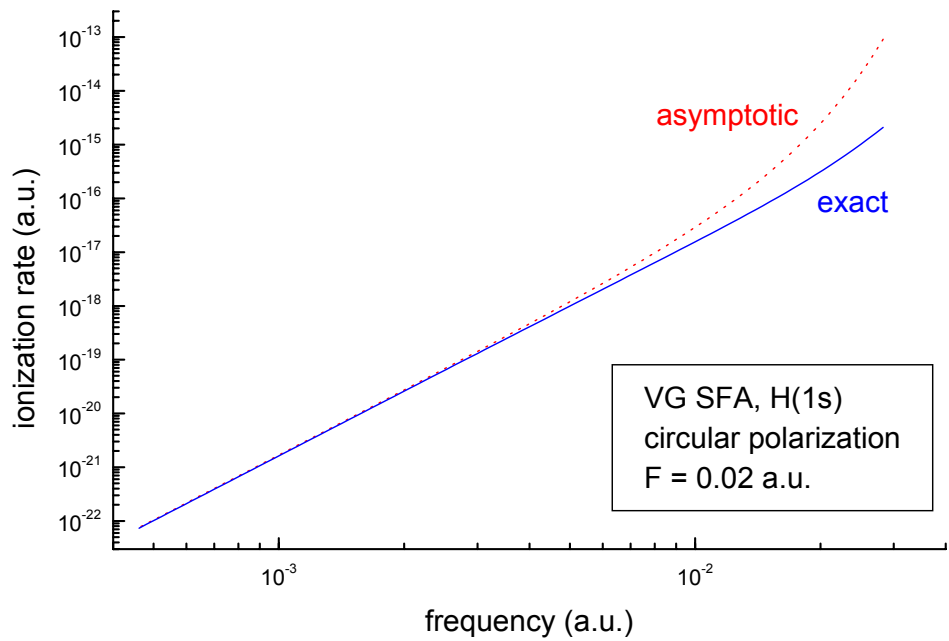

FIG. 3. The VG SFA ionization rates of the $H(1s)$ atom in the circularly polarized laser field for $F = 0.02$ a.u. versus $\omega$.

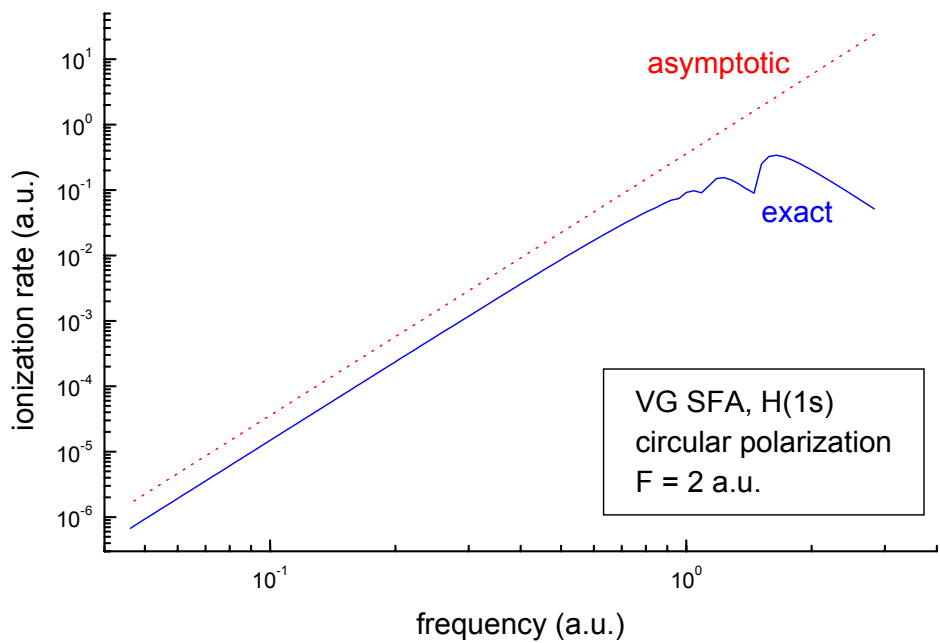

FIG. 4. As Fig. 3, but for $F = 2$ a.u.